\documentclass[english,twocolumn]{aastex63}
\usepackage[T1]{fontenc}
\usepackage[latin1]{inputenc}
\usepackage{subfigure}
\usepackage{amssymb}
\usepackage{float}
\usepackage{graphicx}
\usepackage{amssymb}
\usepackage{bm} 
\usepackage{hyperref}
\usepackage{bm}
\usepackage{amsmath,amsfonts,amsthm,bm} 
\usepackage{babel}

\makeatletter

\newcommand{\av}[1] { \langle {#1} \rangle }
\newcommand{\eq}[1] { Eq.~(\ref{#1})}
\newcommand{\be}{\begin{equation}} 
\newcommand{\ee}{\end{equation}} 
\newcommand{\bea}{\begin{eqnarray}} 
\newcommand{\eea}{\end{eqnarray}}

\makeatother

\received{\today}
\submitjournal{ApJL}

\begin{document}
\title{Discerning between different 'Oumuamua models by 
  optical and infrared observations}

\correspondingauthor{Eirik G. Flekk{\o}y}
\email{e.g.flekkoy@fys.uio.no}

\author[0000-0002-6141-507X]{Eirik G. Flekk{\o}y}
\affiliation{CoE PoreLab, the Njord Centre \\
Department of Physics, University of Oslo, P. O. Box 1048 \\
Blindern, N-0316, Oslo, Norway}
\author{Joachim F. Brodin}
\affiliation{CoE PoreLab, the Njord Centre \\
Department of Physics, University of Oslo, P. O. Box 1048 \\
Blindern, N-0316, Oslo, Norway}
\begin{abstract}
  The first interstellar object to be observed in our solar system
  1I/2017 U1 'Oumuamua combines the lack of observable cometary
  activity with an extra-gravitational acceleration. This has given
  rise to several mutually exclusive explanations based on different
  assumptions in the material composition of 'Oumuamua. We show how a
  combination of  observations in the infrared and optical spectra may
  serve to distinguish between these explanations once another object
  with 'Omuamua-like properties comes close enough to earth. This
  possibility is linked to  the widely different thermal properties of
  the different material models that have been proposed. Developing a
  model for the thermal conduction and infrared signal from a fractal
  model we compare  predictions of the infrared signal
  with that from standard thermal models that assume 'Oumuamua to be
  either a solid piece of rock/ice  or a thin sheet. 
\end{abstract}

\section{Introduction}
 Since the first known interstellar object 1I/2017 U1 ('Oumuamua)
  was discovered in October 2017, much effort has gone into explaining
its formation and unusual behavior (\cite{cuk2018,raymond2018,luu2020,bannister19}).
This behavior is characterized by lack of cometary activity  (\cite{jewitt2017,meech2017}),
a highly elongated shape (\cite{lu2019})
and   a size that is unexpected  from estimated distributions   of small bodies in the solar system or a 
protoplanetary disk  (\cite{jewitt2017, moromartin19a,moromartin2018})-
as well as   a non-gravitational acceleration  (\cite{micheli18}). 
We recently suggested that 'Oumuamua originated as a cosmic
 'dust-bunny',  a cometary fractal aggregate (CFA) that was formed in a cometary tail
 (\cite{luu2020,flekkoy19}). Others  (\cite{cuk2018,fitzsimmons18})
 have proposed it to be a  potentially volatile substance  covered by a rocky crust that was  formed by
tidal disruption and  heating during a close encounter with a nearby star (\cite{cuk2018}).
Another suggestion is that it is a chunk of frozen N$_2$ ejected from an 
exo-Pluto like surface   (\cite{jackson21,desch21}), or a piece of pure
H$_2$  ice  (\cite{seligman20}). Finally, the possibility that it is a
light sail developed by an alien civilization has been advocated
(\cite{bialy18}).
Since 'Omuamua itself is no longer observable, deciding between these
models must await the next passage of a similar object.
Here we show that the combination of optical and infrared observations
offers such a distinction possibility if the passage of the next
object  is as close to earth as was 'Oumuamua.

 We shall refer to these models as the CFA-, ice- rock- and  light sail  
 model. In the case of the CFA and light sail models radiation pressure
 from the sun may account for the non-gravitational acceleration.
 On the other hand, an object of solid  ice or rock is  too massive to be affected by
 radiation pressure, and the non-gravitational acceleration is explained by undetectable outgassing.
The same explanation  has been applied in the case where the sublimating substance is
covered by a rocky crust  (\cite{zhang2020}).

 Infrared observations of 'Oumuamua were limited to those of the  
Spitzer telescope, which had run out of cooling helium (\cite{trilling18}).  
 The new  James Webb telescope to be located at the second Lagrange
 point will offer increased resolution in the 
 infrared spectrum.
  Provided 
 the size and closest distance to earth is comparable to that of 'Oumuamua,
 the combination of optical and infrared observations of another such object would then be sufficient to
 distinguish between the models.  It would require that the optical
observations constrain the shape and rotational state of the object,
 as was the case with 'Oumuamua
 (\cite{jewitt2017, lu2019, jewittluu2019b, mashchenko19}), even though it
had  passed the closest encounter with earth by the time it was first
 observed on UT 2017 October 18.5 (\cite{williams17}).

 In this case the infrared signature would be qualitatively
 different for the different models, since these have different
 thermal properties:
 During observation of the night side  a rock surface will  gradually
 cool. A CFA, on the other hand, is partially transparent to the
 infrared radiation owing to its high thermal conductivity, and will
 gradually heat on the night side. A surface made of N$_2$ ice will stay too cold for detection at all, and, finally, a
 light-sail is  so thin that it has the same infrared
 signature on both sides. 

\section{Thermal models}

As a test-case of these distinction possibilities,  we  take the rotational state and
 observation geometry to be as simple as possible, and use  the known
 values of earth distance, size and shape estimates of  'Oumuamua.

\begin{figure}[h!]
\begin{center}
\includegraphics[width=0.5\columnwidth]{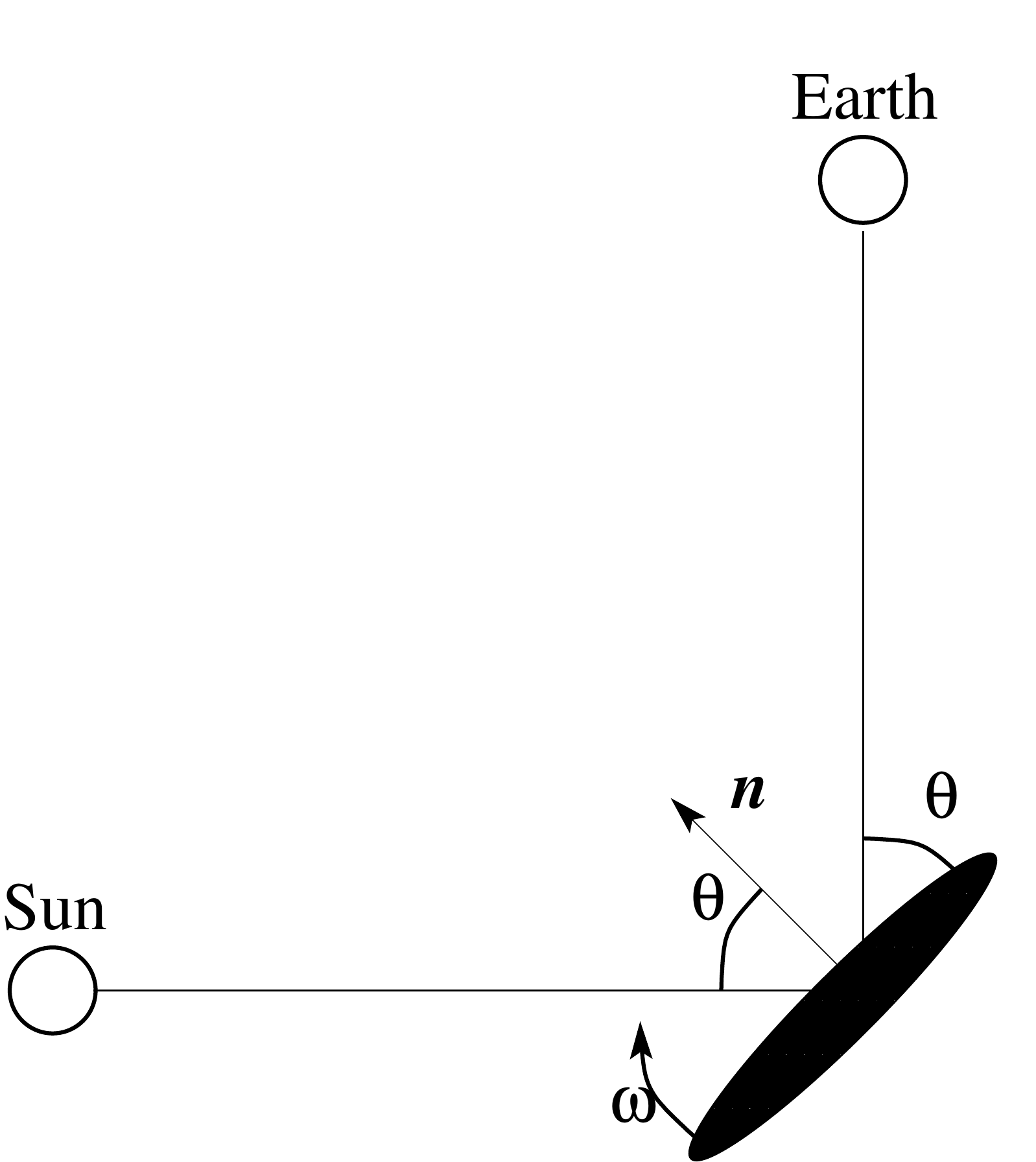}
\caption[]{Top view of the observational geometry showing an oblate ellipsoid
  facing the sun and earth with its broadside during its rotation. The
  phase angle is 90' and $\theta$ the
  angle of incident light to the surface normal $\bf n$. The in-plane rotation is given
by the angular velocity $\omega$.}
\label{geometry}
\end{center}
\end{figure}

In all models the shape is taken to be 
enveloped by an ellipsoid with semi-major axis $a=$ 119 m $b=$ 111 m and
$c=$ 19 m (see Ref. \cite{mashchenko19})  exposing the widest surface area towards the
incoming light during its rotation. The angular velocity $\omega =
2\pi /$7h is that of 'Oumuamua and points in the direction normal to
the plane of figure \ref{geometry}.

In the ice model the absorbed radiation energy from the sun is
consumed by sublimation at a constant sublimation temperature, while
in the other models, it is transported as heat below the surface.
In the   CFA model the dominant mode of this transport is by photons,
while in the light sail and rock model it is by phonons.

The formation scenario of the rock model is suggested to  involve extensive tidal 
fragmentation of a volatile rich parent body during a close H$_2$O ice 
encounter  with its host stars followed by  ejection
(\cite{zhang2020}).
The heating during this 
 process would have created a solid crust of unknown permeability surrounding a more volatile 
 interior. In calculating the surface temperature we shall neglect the
 effect of these volatiles sublimating and only consider the diffusive
 heat transport into a regolith surface.

 Porous materials found in the regoliths of asteroids and comets have
 much smaller values of the thermal conductivity and diffusivity than normal rocks,
 a typical value being $\kappa_t =$ 10$^{-2}$ W/(m K) and 
 $D_t = 10^{-8}$ m$^2$/s (see \cite{cooper03}
 and \cite{jewitt2017}),
 and  even smaller values are assumed  by some authors (\cite{zhang2020}). 
For this reason the  thermal diffusion length in  the rock model is
$\sim$  cm, which is much smaller than the thickness $2c$. 

The surface temperatures of the rock  model is obtained
by thermal modeling (\cite{fitzsimmons18})
based on the  heat diffusion equation
\be
\frac{\partial T}{\partial t }
= D_t  \frac{\partial T}{\partial z^2}
\label{kuyf}
\ee
which  is solved subject to the boundary conditions  
\begin{align}
(1-p)  j_{sun} \cos \theta
  & =  -\kappa_t \frac{\partial T(0,t) }{\partial z }  +  \sigma T^4(0,t) \nonumber \\
  0 & =  -\kappa_t \frac{\partial T(L,t) }{\partial z }
      +  \sigma T^4(L,t)
\label{mhgy}
\end{align}
where $p\sim 0.1$ is the albedo and
$j_{sun}  =$ 1360 W/m$^2$ the solar influx at a distance $R=$ 1AU, $L$
is the thickness in the $z$-direction normal to the surface,
and   $\kappa_t$ is the thermal conductivity. Compared to the standard
thermal model (NEATM (\cite{harris98b})) this description does not
include the 
beaming effect (beaming factor $\eta =$ 1), but does include the
effect of finite thermal inertia.
The thermal diffusivity 
 $D_t = \kappa_t/c_v$, where $c_v$ is the heat capacity per unit volume. 

 The first of the above equations describes the day-side and the last the night  
 side.  
 When $\cos \theta < 0$  boundary conditions
for $z=0$ and $z=L$ are interchanged, reflecting the fact that the day
and night sides are interchanged. The above equations are integrated
using a simple finite-difference scheme and $\theta = \omega t$.

For the frozen $N_2$ the temperature is simply taken to be the
sublimation temperature at zero pressure, $T\approx 63$ K.
In the light sail model the internal diffusive transport of heat may be
neglected and the temperature  assumed to be the same on both sides as
such a sail would have to be much thinner than any reasonable
diffusion length. 
The temperature is then obtained from the energy balance
\be
j_{sun} \cos \theta = 2 \sigma T^4
\ee
where the factor 2 comes from the fact that the sail would radiate
equally on both sides.

\subsection{Thermal conduction in a fractal}
 While the temperature evolution in the ice- sail and rock model is 
described by well established thermal models, the corresponding
transport equations of heat in a fractal structure are less well established.
Diffusion on fractals has been studied extensively
(\cite{procaccia85,havlin02,olsen19}).
In our case however, the transport is not
restricted to the fractal itself, but rather it occurs by radiation in the open
space between the solid sites on the fractal.

In the Methods section it is  show that the mean free path of a photon originating  
from an arbitrary location inside the fractal structure is  
$\overline{\lambda} = (4\pi r_0/3)  
\left( {a}/{r_0} \right)^{3-D}$  where $D=2.35$ (\cite{flekkoy19}) is the fractal dimension and
 $r_0$ the radius of the particles that make up the fractal.
 A photon originating from such a   solid particle, on the other hand,
has a   mean free path  $\sim r_0$, which is  much smaller than $\overline{\lambda}$.  
For this reason  we may
take the radiation field and solid structure to be in local
equilibrium. In other words, they will have the same temperature in
the vicinity of the solid structure.

 On the average, the radiation field then has a constant temperature in
every plane normal to  the surface,  and  there will be  a temperature
gradient in the direction  $\bf n$.
The radiation across a
given plane normal to $\bf n$ 
will be re-absorbed over a distance $\overline{\lambda}$, so that
the net energy flux $j_t$ passing from $z$ to $z+\overline{\lambda}
$ is 
\be
 j_t = \sigma (T^4(z) - T^4(z+\overline{\lambda} ) ) ,
\ee
where $\sigma = 5.67\; 10^{-8}$ W/(m$^2$K$^4$) is the Stefan-Boltzmann
constant.
Taylor expansion of the above expression yields
\be
 j_t \approx  - \sigma \frac{\partial T^4}{\partial z} \overline{\lambda} =  - \kappa_t \frac{\partial T}{\partial z}
\ee
where the thermal conductivity
\be
\kappa_t = \frac{16}{3}r_0 \sigma T^3
\left( \frac{a}{r_0}\right)^{3-D} \approx 0.8 \mbox{W}/(\mbox{K m})
\ee
if the values  $r_0=1\mu m$ and $T=250$ K  are
used (\cite{flekkoy19}). This value is in the range of the thermal conductivity $\kappa_0
\sim$ 1 W/(K m)  for silicate rocks. 

On the other hand, the thermal conductivity $\kappa_s$ of the solid
structure that makes up the fractal
depends on the solid fraction $\phi_s = (r_0/a)^{3-D} \sim 10^{-5}
$(\cite{flekkoy19}), through the relation
$\kappa_s = \phi_s \kappa_0 \sim 10^{-5}$ W/(Km). This means that the thermal conduction
of the radiation field is $\sim$ 5 orders of magnitude larger than that
of the solid. 

The heat capacity per unit volume  of the radiation
field is given as $c_v = \partial \epsilon /\partial T$ where the
energy density $\epsilon = \sigma T^4/c_0$,  and $c_0$ is the speed of light.
It may be written
\be
c_v = \frac{1}{c_0}
\frac{\partial j}{\partial T} = \frac{4\sigma T^3}{c_0} \approx 1.2\; 10^{-8} \mbox{J/(m$^3$K)}
\ee
where $j= \sigma T^4$.
This value is 8-9 orders of magnitude smaller than the average heat capacity of the
solid, which is given as
\be
c_{vs} = c_{v0} \phi_s \approx 5  \mbox{ J/(m$^3$K)}
\ee
where $c_{v0} \sim 10^6 $J/(m$^3$K) is the typical heat capacity of
rocks. So, while the radiation field governs the heat conductivity, the
solid phase governs the heat capacity, as was also found by Merril
(\cite{merril69}) who studied heat transfer in evacuated powders.
As a result the thermal diffusivity
\be
D_t = \frac{\kappa_t}{c_{vs}} \approx 0.14  \mbox{ m$^2$/s} .
\label{Dtcfa}
\ee
The corresponding thermal diffusion length in the CFA  over a time $t=3.5$ h
(the half period of 'Oumuamua)  $x_t = \sqrt{2D_tt} \approx$ 60 m,
which is significantly larger than the estimated thickness $2c=$ 38 
m (\cite{mashchenko19}).

\begin{figure}[h!]
\begin{center}
\includegraphics[width=0.50\columnwidth]{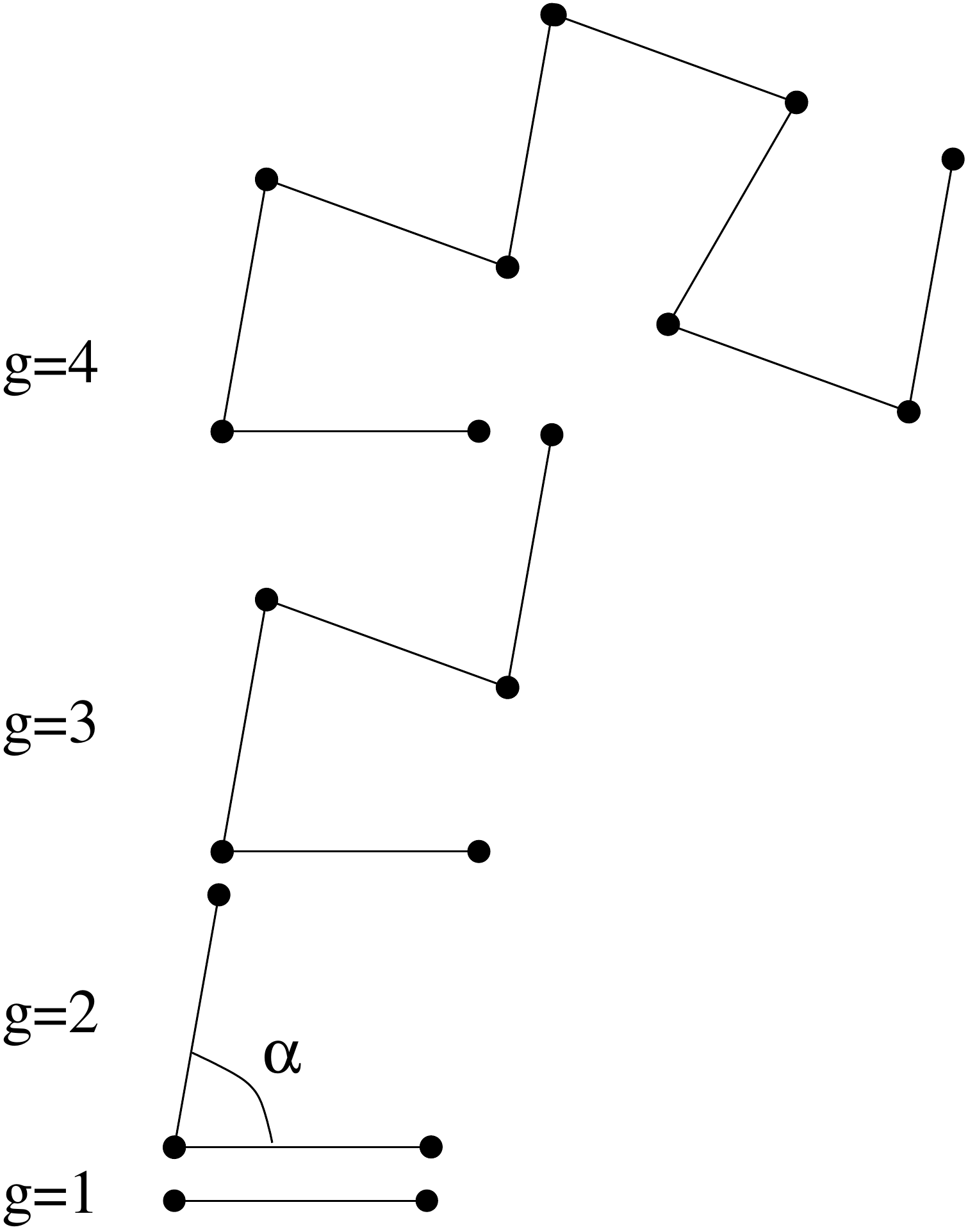}
\caption[]{The first 4 generations of the fractal model structure.}
\label{figfrac}
\end{center}
\end{figure}

This implies a   transparency to infrared radiation, which will vary
with location.
 Since any fractal structure has inhomogeneities on all length scales,
 geometric fluctuations will cause temperature variations on all scales as
 well.  For the purpose of quantifying the effects of these
 fluctuations we construct a fractal of the prescribed dimension
 $D=2.35$.
 It   is
constructed by a hierarchical procedure which is illustrated in figure
\ref{figfrac}.   We start with two points at a unit separation in a plane
with coordinates $x$ and $y$. Then, at every generation $g$, 
a copy of the entire structure is rotated an angle $\alpha$ around the
end point. 

The consecutive rotations illustrated in figure \ref{figfrac}
produces an
ordered structure that is confined to the $xy$-plane. 
 In order to introduce randomness as well as a structure
that extends in 3 dimensions, two additional steps are added to the
model: First, the replacement $\alpha \rightarrow \pm \alpha + \delta
\alpha$, where $\delta \alpha$ is a random addition of zero mean and
$\sqrt{\av{\delta \alpha^2}} = \alpha /50$, is  carried out. 
Second, an out-of-plane tilting  by an average angle of $1/6$ is
performed. This has the effect of giving the overall structure an
envelope of aspect ratio $c/a= 1/6$ as indicated by the fitting of the
'Oumuamua light curves (\cite{mashchenko19}).

It should be noted that the present
 model does not represent the physics of the aggregation processes
 leading to fractal structures (\cite{suyama08,katoaka13,okuzumi12,wada11}), but only
 seeks to capture the geometric fluctuations that are intrinsic to
 such fractals. It does, however, mimic the buckling process that is
 caused by colliding dust aggregates (\cite{suyama08}),  by prescribing an
 angle between connected particle chains, the smaller the angle, the
 larger the fractal dimension. To get the prescribed $D=$2.35 value an angle of
 $\alpha = 0.48 \pi$ was used (see appendix). 
 
\begin{figure}[h!]
\begin{center}
\includegraphics[width=0.6\columnwidth]{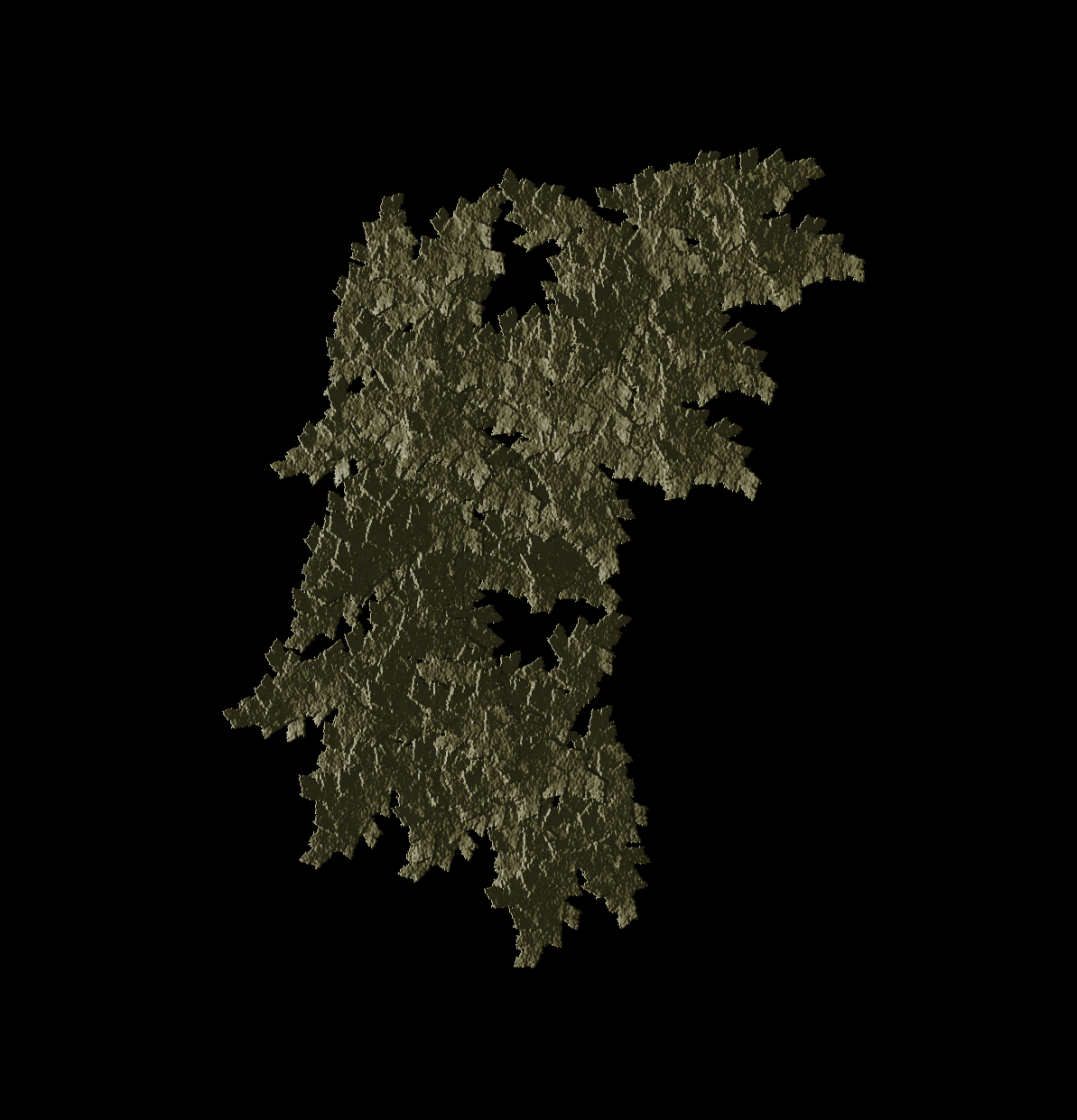}
\includegraphics[width=0.6\columnwidth]{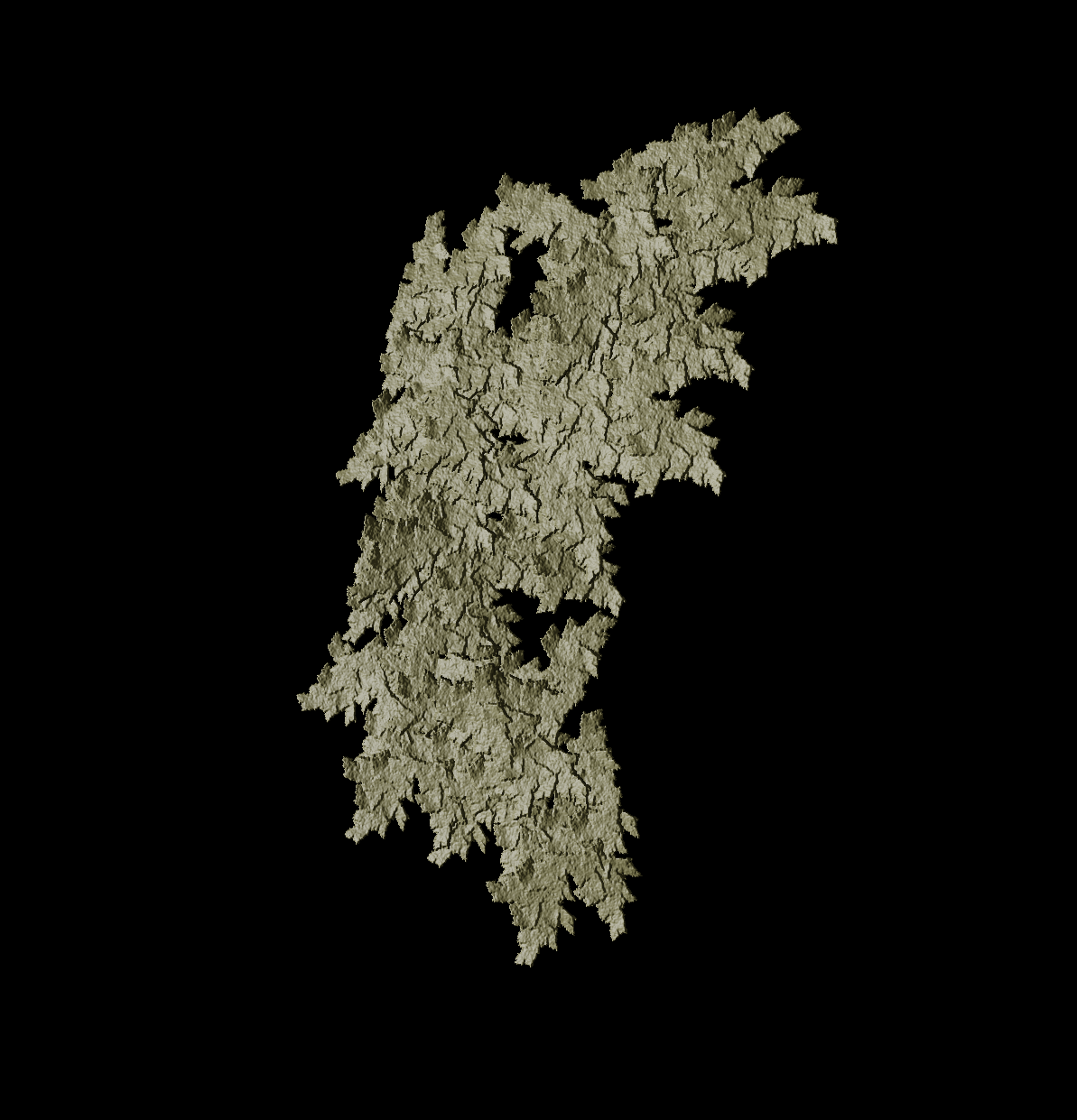}
\includegraphics[width=0.6\columnwidth]{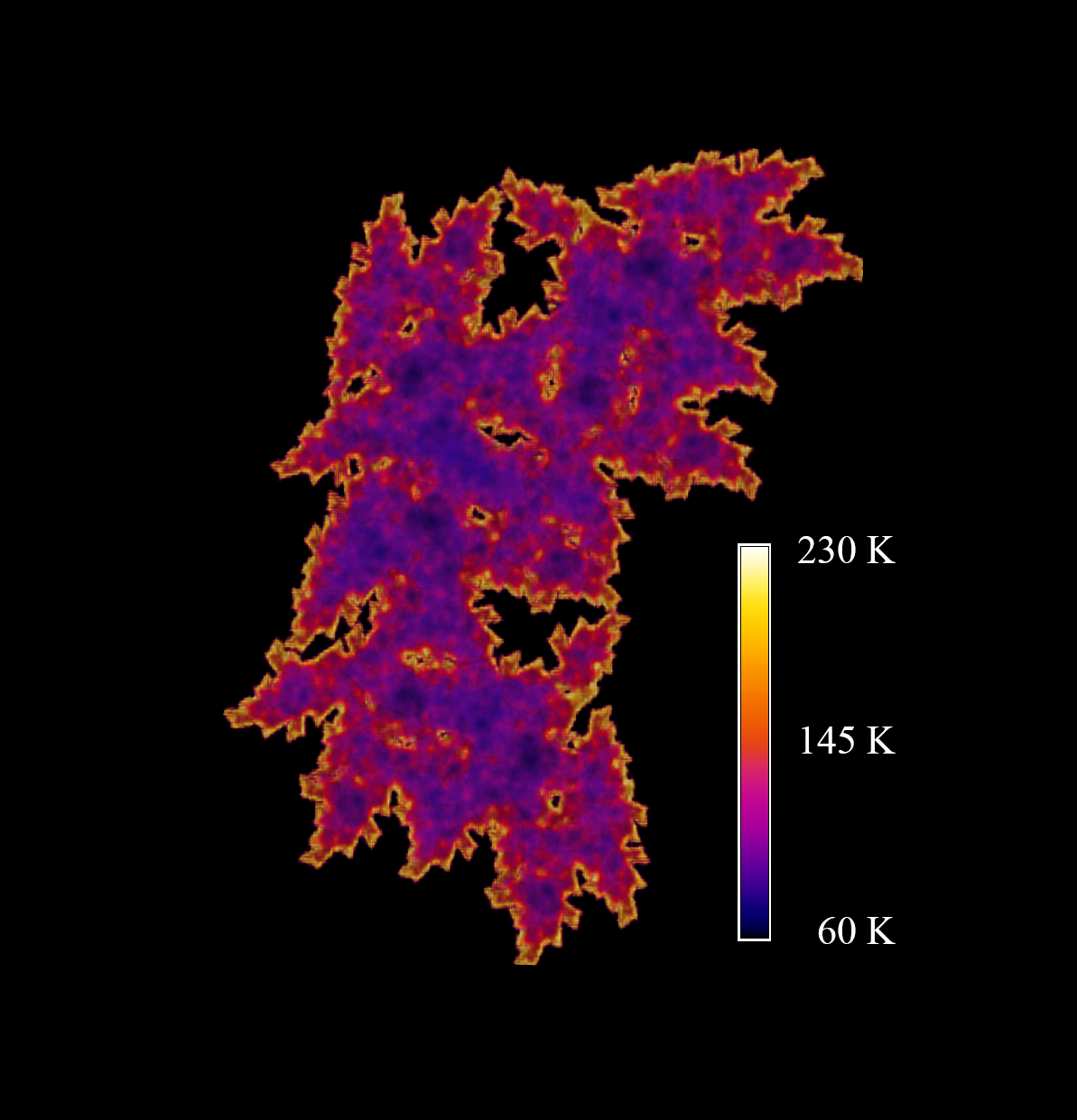}
\caption[]{Optical day side- and infrared night side (bottom figure) radiation from
  the geometric   CFA model of 'Oumuamua using 8 million particles.
  The day side images correspond to 
  $\theta=$ 0' and 30' in figure \ref{geometry}, while the infrared 
  image is a nightside view of the $\theta=$ 0' orientation.}
\label{fig1}
\end{center}
\end{figure}

The fractal model is applied to represent the local thickness
fluctuations in the  $z$-direction:
The $xy$-projection of particle  density $\rho (x,y)$  defines the  local thickness
\be
L(x,y)  = \frac{\rho(x,y)}{\overline{\rho}} 2c
\ee
where $\overline{\rho}$ is the average of $\rho (x.y)$, so that the
average of $L$ is $2c$.  This local $L$-value is then taken as input in \eq{mhgy} to obtain
the local day- and night side temperature shown in figure \ref{fig1}, which
also shows two optical images. Note that regions of high infrared transparency exist on all 
scales. 

\begin{figure}[h!]
\begin{center}
\includegraphics[width=0.9\columnwidth]{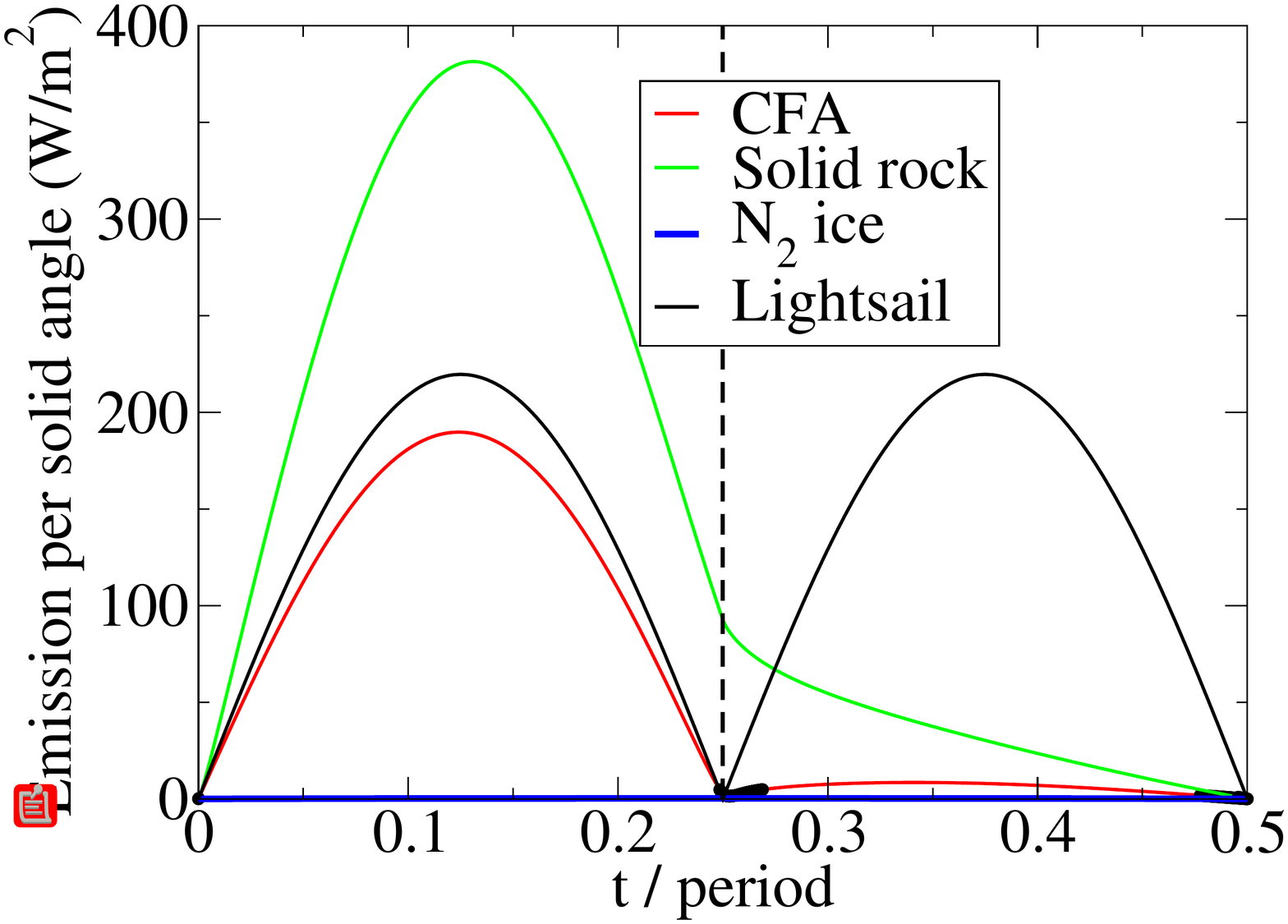}
\includegraphics[width=0.9\columnwidth]{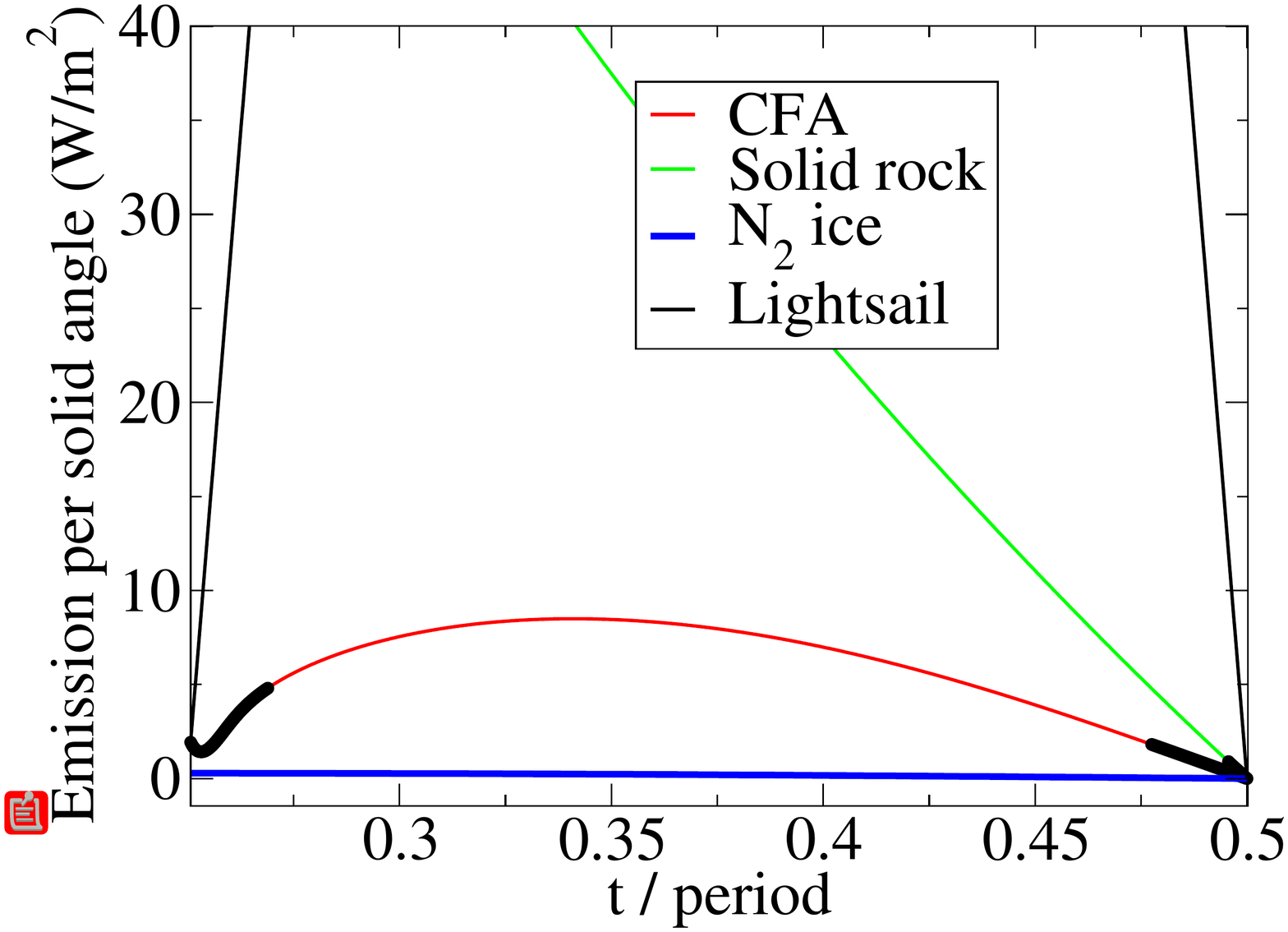}
\caption[]{Top figure: Infrared signals in the direction of earth from the different models over half a
  period of rotation. The vertical line shows the transition from the
  day- to night side, and the thick black line mark where the signal falls
  below the sensitivity levels of MIRI. The observational distance is
  0.16 AU. Bottom figure: Zoom-in of the night side infrared signal.}
\label{comparison}
\end{center}
\end{figure}

However, observations by an infrared telescope are unlikely to resolve
the level of detail shown in this figure.
In order to determine the average effect of the geometric fluctuations
inherent in a fractal we may simply integrate the radiation over the
$xy$-plane. Taking the fluctuations into account in this way we may define 
the  effective thermal thickness $L_{eff}$
that gives the same radiation
from a disc with constant thickness (see the appendix).
 In the limit of large system sizes an asymptotic value of $L_{eff}$  is expected from the fractal
nature of the geometry.
In the Methods section we obtain the value $L_{eff} \approx 2c/5$, so,
the fluctuation effect is large; it reflects the  non-linear
relationship between $T$ and $L$. 
 Using the $L\rightarrow L_{eff}$ replacement in \eq{mhgy} allows  for a one dimensional calculation of the 
radiation at each moment in  time as $\theta$ increases. This was done
calculating the infrared light curves in figure \ref{comparison}.

\subsection{Condition for observations } 
The reported sensitivity of the MIRI imager of the James-Webb telescope gives the signal
strength at which the signal-to-noise ratio is 10 \footnote{See figure
  1 in user
  documentation for the James Webb Space Telescope, MIRI sensitivity  ( https://jwst-docs.stsci.edu/jwst-mid-infrared-instrument/miri-predicted-performance/miri-sensitivity). }
for an  on-source integration time of 10 ks. Reducing the integration time to
1 ks allows for the resolution of time variations on the time scale of
'Oumuamuas rotation period. 
This implies  a corresponding increase in the noise floor by a factor
of 10. In this case   curve  fitting  of the predicted
MIRI  noise floor $s$, yields the  approximation
      \be 
      s = 10^{ -32.0 + 2(\lambda- \lambda_0 )/\Delta \lambda }
      \mbox{  W/(m$^2$ Hz)}
\label{sensitivity}
      \ee 
      where $\lambda $ is the wavelength of the observed radiation
    $\lambda_0 = 4.0$ $ \mu$m and      $\Delta \lambda =
      18.5 $ $\mu$m.
To get the comparable prediction of the signal strength we use the wavelength of maximum
intensity $\lambda_m(T) = b/T$, where   $b = 2.9$ $10^{-3}$ K m is the
constant of Wiens displacement law. The standard Planck spectrum then  gives 
      \be
      I_m (T)= 2 \frac{|\sin( \theta (t) )| h\nu_m^3(T)}{c_0^2(e^{hc/(kb)}-1)}
      \left( \frac{ a}{R_o} \right)^2
\label{strength}
      \ee
where $\nu_m (T)= c/\lambda_m(T) $, $h$ is Plancks constant, $c_0$ the speed of
light and $R_o=$ 0.16 AU is the observation distance.

\subsection{Predicted infrared signals for different models} 
Figure \ref{comparison} shows the result for all 4 models over half a
period, beyond which all curves repeat themselves. All thermal models
are run for a number of initial rotations until their light curves
have converged to steady state values. Only the rock model maintains an
internal temperature below the diffusion skin depth. However, changing
this internal temperature only changes the surface temperature by an
amount  $\sim$ 1 K, and the steady state values are  reached to within $\sim$ 1\%
by 3 rotation periods.

Except for the ice model signal, which falls below the detection level at all times, the
signals are masked by a black line wherever they fall below the
detection level. This level is defined by the temperature where
$I_m(T) < s$, where  the sensitivity $s$ is defined in
\eq{sensitivity} and  $I_m$ the predicted intensity
given in \eq{strength}.

Most notably, while the rock model displays a steady signal decay in
over the night side period, the CFA model produces a second observable
maximum.
This is the case for the lightsail model as well, but this signal is
easily recognizable since the day- and night side maxima have the
same values.  Also, only the rock model with its significant heat
capacity has a detectable signal at $ t=$  period /4, at which point
the CFA becomes invisible in the infrared region. 

\section{Discussion }

Having identified a set of crucial measurements that distinguish between the
different hypothesis for the structure of the next 'Oumuamua object using
the existing James-Webb telescope has clear advantages. 
Technical solutions for  chasing it with a dedicated spacecraft that
could make close observations, have 
been suggested (\cite{hibberd20,seligman18}),  and will be much more challenging.

Analysis of 'Oumuamua  light curves indicates a tumbling rotational
state (\cite{drahus2018,fraser2018}), and the different 'Oumuamua models are all likely to result in a such a
 state. The models that rely on the radiation pressure to
explain the extra-gravitational acceleration would likely acquire such a
state from the YORP-effect (\cite{rubincam2000}).  In the rock models where the acceleration is caused by out-gassing, 
 a tumbling state would likely result from
the torque created by the gas pressures (\cite{rafikov18}). Also, in the case of the rock model
a tumbling state may have survived the inter-stellar travel (see
Ref. (\cite{burns73}) from lack of internal dissipation caused by
rotational deformation.
Such tumbling has not been included in our calculations. Yet, these
calculations
show that the infrared signals from the different 'Oumuamua models will be
qualitatively different.

 Different shapes and rotational states will affect both the infrared
 and optical signals. In the case of the oblate ellipsoidal shape, which emerged as the more
likely one for 'Oumuamua
(\cite{mashchenko19}), there is significant rotation around a minor
principal axis (the major principal axis being associated with the
maximum moment of inertia). This explains the large light-curve
variations, since rotation purely around the major principle axis
would cause no light curve variations at all. 

If the case of a prolate shape, rapid rotation around a minor principal axis could make the corresponding rotation period shorter than the thermal relaxation time. 
This would blur out the infrared signal variations since the
temperature would then even out on the different sides, thus making
the signals from the CFA and rock models similar.   However, this
behavior would be predictable from a proper inversion of the
light-curve data with respect to the rotational state. So, the cases where the infrared signal is less effective as a tool to discern the different models, are identifiable. 

Optical observations that  constrain the shape and rotational state
of the object  (\cite{mashchenko19}), will therefore make it possible to obtain
correspondingly different predictions for the infrared signal of the
different models, thus making it possible to distinguish between
them.  The main difference between the infrared predictions for the
different  models is most pronounced in the night-side signal where
the CFA model produces a weaker  maximum that is not present in the other
models. Since this maximum is only a factor 2 above the 10 SNR noise
floor at an observation distance of 0.16 AU (the closest approach of
'Oumuamua),  the distinction possibility is limited to  near-earth observations.

\appendix
\label{app1}

\subsubsection*{Fractal model}
The fractal dimension is obtained by noting that the
overall size of the structure is increased by a factor $(2 - 2\cos
\alpha)$ as $g \rightarrow  g+1$, while the number of links in the
structure increases by a factor 2.  At generation number $g$ the 
total size of the structure $L_g =  (2-2\cos \alpha)^g$ and the number of
links $M_g=2^g$.  
Eliminating $g$ between these two equations yields $M=L^D$ where the
fractal dimension 
\be
D= \frac{\ln 4}{\ln (2 - 2 \cos \alpha)},
\ee
or, equivalently
$\cos \alpha =1 - 2^{(2-D)/D} $.
\begin{figure}[h!]
\begin{center}
\includegraphics[width=0.44\columnwidth]{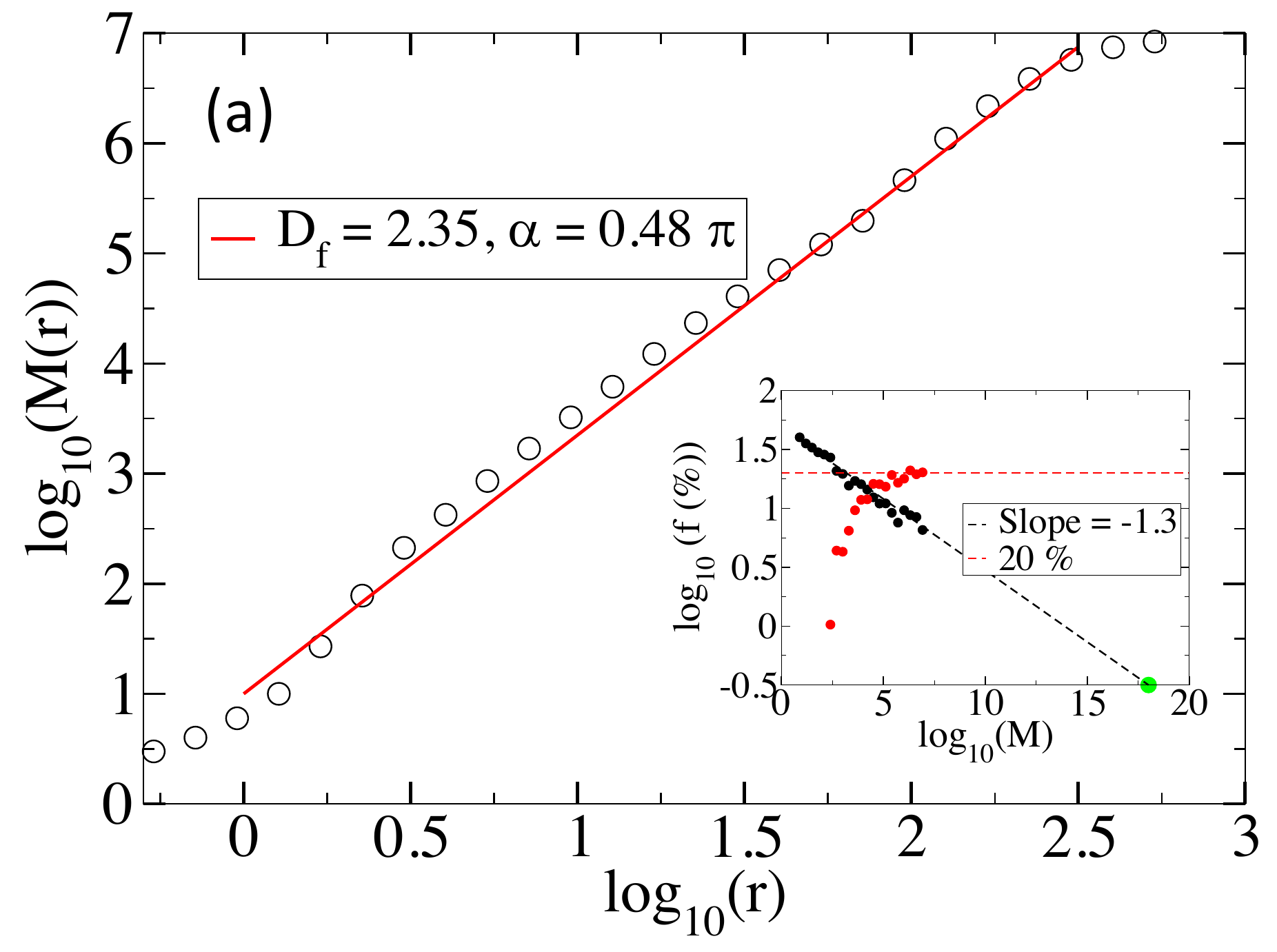}
\includegraphics[width=0.47\columnwidth]{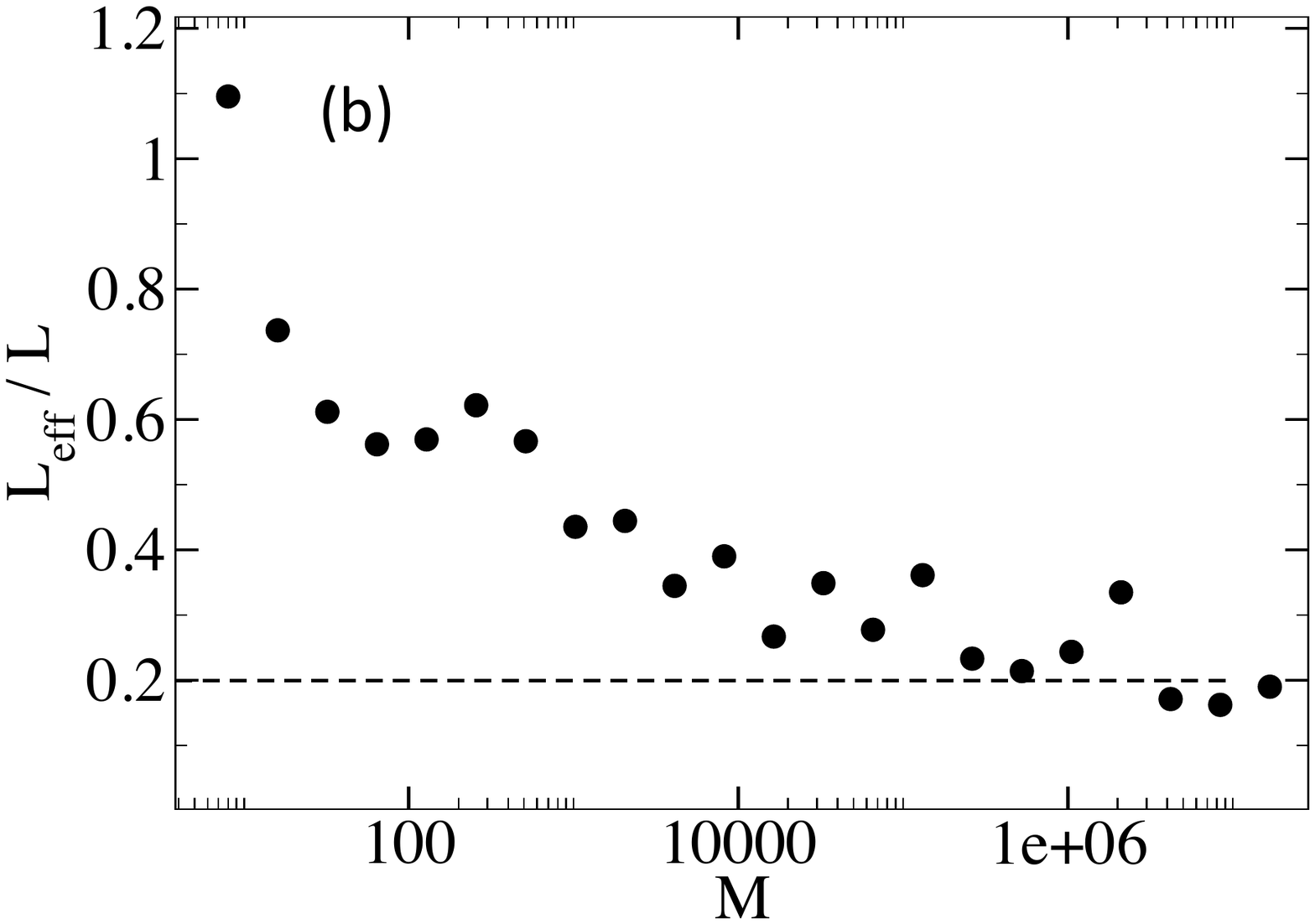}
\caption[]{(a) The number of particles (mass) of the fractal structure as
  a function of distance from the initial $g=1$ structure. The red
  line has slope 2.35. The insert shows 
results from  projections of the fractal particle structure 
  into the $xy$-plane.    Black dots show the fraction of sites $f$, that contain only one 
  particle, red dots the fraction of sites that contain less than 10 
  \% of the average particle number. The green dot shows the 
extrapolation to the  number of particles in an 'Oumuamua size
fractal. (b) The average effective thermal thickness as a function of system size.}
\label{Mr}
\end{center}
\end{figure}

Figure \ref{Mr} (a)  plots  the particle number as a function of
distance from the $g=1$ starting position for a system of 8 million particles.
It shows that the behavior is
indeed fractal over 2-3 orders of magnitude. The cross-over behavior
at large scales happens as $r$ approaches the system size.

The number  of particles in  an object like 'Oumuamua is about 8
orders of magnitude larger than in our simulations; a simulation of
such  particle number is  beyond the capacity of any existing computer.
However, relative numbers, such as the volume fraction of regions that
have a certain fraction of the average density, will be constant in
systems large enough to avoid significant finite size effects.
The general reason for this is that a crossover at a certain scale
would  define a length scale which is different from both the system
size and the particle size, and the defining feature of a fractal is
exactly that it lacks such intermediate scales.

In the insert of figure \ref{Mr} (a)   the fraction of sites that contain
less than  10 \% of the average particle number after projection into
the plane that contains the largest semi-major axis $a$ and $b$ is plotted.
The particular number 10 \% is chosen arbitrarily to 
define regions that are significantly thinner than the average, the point
being  that  this fraction  approaches an asymptotic value
already at a particle number of a million. Below that number the
average number of projected particles is sometimes below 10, which
makes it impossible for a non-zero particle number to be below 10 \%
. This is a finite size effect.
Figure \ref{Mr} (a)  also plots  the fraction of projected
particle numbers that equal  one. This fraction decays as a power-law
with an extrapolation to a few parts per thousand at the size of
'Oumuamua ($M\sim 10^{18}$), which shows that only a very small
fraction of the  fractal will contains sites that emit
infrared radiation without further scattering.
This further justifies the assumption of a local equilibrium between
massive particles and the radiation.
\subsubsection*{Photon mean free paths in a fractal}
 In the following we derive the mean free paths $\lambda$ of a photon
emitted from a solid site on a fractal of dimension $D$ and the mean
free path of one starting from an arbitrary location, starting with
the former.
In order to do this we consider first the probability $P_0(r)$ of an
emitted photon not hitting
another particle of radius $r_0$ over a distance $r$.  If we split this distance into
$n$ segments, each of length $\Delta r = r/n$,  we can write $P_0(r)$
as the product of the probabilities of not hitting a particle in each
of these segments:
\be
P_0(r) = \prod_{i=0}^n (1- \pi r_0^2 \rho_N(r_i) \Delta r) .
\label{product}
\ee
Here $r_i = i\Delta r$ and $\rho_N(r)$ is the number density of
particles so that $\Delta r \pi r_0^2 \rho_N(r_i) $  is the average
number of particles in the volume  $\Delta V=\Delta r \pi r_0^2 $. This
small average particle number equals the probability of  the photon stopping inside
this volume, and consequently, the expression in the parenthesis in
\eq{product}  is the probability of not hitting a particle inside
$\Delta V$.

Now, taking the log of \eq{product}  gives
\begin{align}
\ln P_0(r) & = \sum_i \ln (1- \pi r_0^2 \rho_N(r_i) \Delta r)
             \nonumber \\
  & \approx -\sum_i  \pi r_0^2 \rho_N(r_i) \Delta r 
             \nonumber \\
  & \approx -\int_{r_0}^r dr'  \pi r_0^2 \rho_N(r') 
\end{align}
where we have used the approximation $\ln(1-x) \approx -x$ for $x\ll
1$ in going
from the first to the second line, and taken the $\Delta r \rightarrow
0$ limit in passing to the last line. Exponentiating gives
\be
P_0(r) = e^{-\int_{r_0}^r dr'  \pi r_0^2 \rho_N(r') } .
\label{P0}
\ee
Over the distance  
$r$  the photon will either be absorbed or not.  So, the
probability of being absorbed between $r_0$ and $r$ is therefore $P(r)
= 1 - P_{0}(r)$. The probability $p(r) dr$ of being absorbed between $r$ and
$r+dr$ is therefore
\be
p(r) dr = P(r+dr) - P(r) = P'(r) dr .
\label{der}
\ee
Note that $p(r)$ is also the distribution of the mean free paths,
which we can now write
\be
p(r) = \pi r_0^2 \rho_N(r) e^{-\int_{r_0}^r dr'  \pi r_0^2 \rho_N(r') } 
\label{kug}
\ee
by using \eq{P0} and \eq{der}.  The mean free path is then given by
\be
\lambda = \int_{r_0}^{\infty} dr r p(r) . 
\ee
Using the fact that the number density around a particle in a fractal
of dimension $D$ is (\cite{flekkoy19})
\be
\rho_N (r) = \frac{3}{4\pi r_0^3}
\left( \frac{r}{r_0} \right)^{D-3}
\ee
the exponent in  \eq{kug} becomes
\be
\int_{r_0}^r dr'  \pi r_0^2 \rho_N(r')  = -\frac{3}{4(D-2)}
\left( \left(
\frac{r}{r_0}  \right)^{D-2} -1  \right) 
\ee
and with the substitution $x=r/r_0$, we find the mean free path
\be
\lambda =  r_0 \int_1^{\infty} dx x^{D-2}e^{-\frac{3}{4(D-2)}
  (x^{D-2}-1)},
\ee
which  we can write as $\lambda = I(D) r_0$, where the prefactor
$I(D)\sim 1$ as long as $D>2$. When $D<2$, however, $I=\infty$, and 
 $I\rightarrow \infty$ as $D\rightarrow 2^+$. For $D=2.35$,
 as in our case, $\lambda \sim r_0 $.

The  probability $P_0(a)$ that a
photon originating inside the structure will escape over a distance
$\sim a$. is given by \eq{P0}, which gives
\be
P_0(a) = e^{-\frac{3}{4(D-2)}\left( \left(
\frac{a}{r_0}  \right)^{D-2} -1  \right) 
}
\ee
which is extremely close to zero as $(a/r_0)^{D-2} \sim 10^5$.

The mean free path of a photon starting from an arbitrary point on a surface
that cuts through the  fractal is not determined by the mass density surrounding
a solid point, but rather the average density on that surface. 
This surface, as well as cross-sections parallel to it, will have an
average number density of particles
\be
\rho_N(a) = \frac{3}{4\pi r_0^3}\left( \frac{a}{r_0} \right)^{D-3}.
\label{jytdfu}
\ee
In order to estimate the mean free path $\overline{\lambda}
$ from such a surface we require
that the volume $\overline{\lambda} \pi r_0^2$ be equal to the average
volume per particle $1/\rho_N(a)$.  This gives
\be
\overline{\lambda} = \frac{1}{\pi r_0^2 \rho_N(a)}= 
\frac{4\pi r_0}{3}
\left( \frac{a}{r_0} \right)^{3-D}. 
\label{jytdf}
\ee

\subsubsection*{Fractal fluctuations and the effective heat thickness $L_{eff}$}
The thickness fluctuations of the fractal CFA model will cause local
temperature fluctuations. By averaging the corresponding radiation
$\sigma T^4$ from the surface, it is possible to define an effective thickness
$L_{eff}$ that produces the same radiation from a disc of that thickness.
For the CFA model, where the thermal diffusion length exceeds the
thickness $2c$, a steady state assumption is justified, in which case
we may replace the temperature gradients in the boundary conditions
by the approximation
\be
\frac{\partial T(L,t) }{\partial z }  \approx \frac{Tb-T_f}{L(x,y)}
\label{Tapprox}
\ee
where $T_f$ and $T_b$ are the front- and backside surface temperatures.
Then the steady state is described by the energy balance
\begin{align}
(1-p) j_{sun} \cos \theta (t) & = \sigma T_f^4 - \kappa_t
                                \frac{T_f-T_b}{L(x,y)}
                                \nonumber \\
\kappa_t \frac{T_f-T_b}{L(x,y)} & = \sigma T_b^4
\label{balance}
\end{align}
where  $T_f$ is the day
side temperature and $T_b$ the night side temperature. Indeed, solving
the full diffusion equation (\ref{kuyf}) with the CFA parameters give
temperature profiles $T(x,t)$ that   are quite linear in $x$,
justifying the use of \eq{balance} in calculating $L_{eff}$. 
Figure \ref{Mr} (b) shows how this effective thickness varies with 
system size. In these calculations $L_{eff}$ was averaged over 20 
different structures for each system mass $M$. 
The fact that it converges to an approximate asymptotic value 
indicates that the simulations are in the proper  
large-size regime. 

So, equations (\ref{balance}) were used as an approximation to produce  figures \ref{fig1}.
In the calculations   produced
the temperature values in figure \ref{comparison}
the substitution  $L\rightarrow L_{eff}$ was used in \eq{mhgy}  to 
represent the net effect of local variability  in the transmission of heat.

\subsubsection*{Acknowledgments }
We thank
Jane X. Luu and Renaud Toussaint for early discussions on this work, as well as 
the Research Council of Norway through its Centers of Excellence funding scheme, project number 262644.


\end{document}